# Origin of the injection-dependent emission blueshift and linewidth broadening of III-nitride light-emitting diodes

Nick Pant[1,2], Xuefeng Li[3], Elizabeth DeJong[3], Daniel Feezell[3], Rob Armitage[4], and Emmanouil Kioupakis[2]

[1]Applied Physics Program, University of Michigan, Ann Arbor, MI 48109, USA
[2]Department of Materials Science & Engineering, University of Michigan, Ann Arbor, MI 48109, USA
[3]Center for High Technology Materials, University of New Mexico, Albuquerque, NM 87106, USA
[4]Lumileds LLC, San Jose, CA 95131, USA

Corresponding author: nickpant@umich.edu

**Abstract**
III-nitride light-emitting diodes (LEDs) exhibit an injection-dependent emission blueshift and linewidth broadening that is severely detrimental to their color purity. Using first-principles multi-scale modelling that accurately captures the competition between polarization-charge screening, phase-space filling, and many-body plasma renormalization, we explain the current-dependent spectral characteristics of polar III-nitride LEDs fabricated with state-of-the-art quantum wells. Our analysis uncovers a fundamental connection between carrier dynamics and the injection-dependent spectral characteristics of light-emitting materials. For example, polar III-nitride LEDs offer poor control over their injection-dependent color purity due to their poor hole transport and slow carrier recombination dynamics, which forces them to operate at or near degenerate carrier densities. Designs that accelerate carrier recombination and transport and reduce the carrier density required to operate LEDs at a given current density lessen their injection-dependent wavelength shift and linewidth broadening.

Although III-nitride light-emitting diodes (LEDs) have been highly successful for producing blue light efficiently, they face several challenges for the longer green and red wavelengths.[1] Their wall-plug efficiency decreases as the emission wavelength increases and becomes worse for high-power operation, a phenomenon known as the green gap.[2–6] Another challenge is the blueshift of the emission wavelength and the broadening of the spectral linewidth with increasing carrier injection. These effects change the perceived hue, which severely deteriorates the color purity of LEDs at high operating powers.[7] In many cases, the perceived hue is blueshifted, and this worsens the efficiency gap by requiring even longer wavelength devices to compensate for the perceived blueshift. Despite the overwhelming technological importance of this problem, a quantitative understanding of the injection-dependent spectral blueshift and linewidth broadening has been missing.

The band-edge emission of polar InGaN quantum wells is determined by the interplay of competing mechanisms that contribute to the emission by shifting the band gap or by filling the bands (Figure 1). To date, the most widely accepted explanation of the injection-dependent blueshift is screening of polarization fields by free carriers, with a smaller role attributed to phase-space filling.[8,9] Meanwhile, there is no widely accepted explanation for the origin of the linewidth broadening. III-nitride quantum wells exhibit strong piezoelectric and spontaneous polarization fields, which contribute to a quantum-confined Stark shift of the band gap.[10–12] As free carriers are injected into the quantum well, they screen the polarization charges, which results in a blueshift of the band gap as the bands flatten (Figure 1(a)). A competing, and often overlooked, effect that redshifts the energy is the renormalization of the band gap by many-body effects in the free-carrier plasma,[13–16] an effect that has been directly measured in bulk samples.[17,18] At carrier densities exceeding $10^{18}$ cm$^{-3}$ relevant for LED operation, excited carriers exist predominantly in the correlated plasma state rather than as bound excitons,[19] due to Pauli blocking and screening of the Coulomb interaction.[20] An electron (hole) in a plasma repels other electrons (holes), creating a surrounding region of positive (negative) charge, called the exchange-correlation hole.[21] The net result is an effective attractive potential for the carriers, which lowers the conduction band and raises the valence band as the carrier density increases (Figure 1(b)). In contrast to band-gap shift effects, phase-space filling contributes to a blueshift of the peak-emission energy by changing the occupancies of the bands.[10–12] As the carrier density increases and the quasi-Fermi levels penetrate deeper into the bands (Figure 1(c)), the emission occurs from states that are further away from the band edge. This effect becomes pronounced only if both carriers are degenerate. Therefore, the emission of InGaN quantum wells is influenced by the complex interplay of band-gap shift and band-filling effects in the free-carrier plasma.

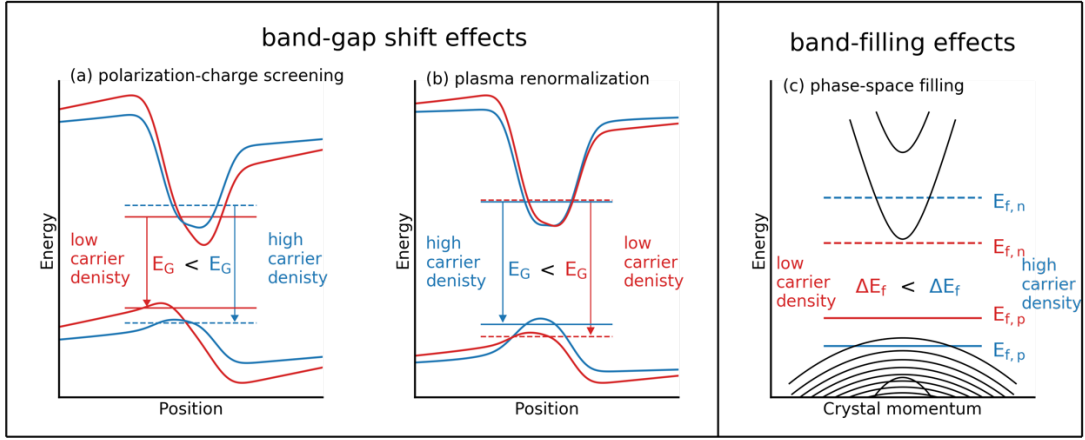

Figure 1. Schematic illustrations of the three primary effects that contribute to the band-edge emission of polar III-nitride quantum wells at carrier densities relevant for LED operation. Band-gap shift effects such as polarization-charge screening (panel (a)) and plasma renormalization (panel (b)) contribute to the emission spectrum by shifting the band gap $E_G$. Band-filling effects such as phase-space filling (panel (c)) contribute to the emission spectrum by changing the finite occupation of carriers (indicated in the figure by the electron and hole quasi-Fermi levels $E_{f,n}$ and $E_{f,p}$, and their difference $\Delta E_f$), which in turn determines the region of phase-space from which carriers recombine to produce light.

An experimental understanding of the band-edge emission of InGaN LEDs has been impeded by the difficulty in distinguishing the competing effects. For example, Kuokstis et al. compared the luminescence of bulk films against quantum wells to isolate the effects of phase-space filling from polarization-charge screening.[22] However, this approach assumes that polarization fields do not affect phase-space filling, which is not true as we will show later. On the other hand, several experimental works have attempted to explain the injection-dependent broadening of the high-energy tail of the luminescence spectrum in terms of carrier delocalization.[23–27] Although these works reveal interesting correlations, it is difficult to establish causation from their data. On the theoretical front, previous studies have not explained the experimentally observed injection dependence of the blueshift and linewidth broadening. Della Sala et al. used self-consistent tight-binding simulations in the virtual-crystal approximation to conclude that polarization-charge screening is responsible for the injection-dependent blueshift but they neglected phase-space filling, carrier localization, and many-body renormalization.[28] On the other hand, Peng et al.[29] neglected alloy disorder, and it is unclear what simulation parameters they used to match experimental data since the work dates from a time when various fundamental parameters, e.g., the band gap of InN[30] and polarization constants,[31] were not accurately known. Therefore, a theory of the injection dependence of the emission spectrum of III-nitride LEDs is entirely missing.

In this work, we use first-principles multi-scale modelling to explain the carrier-injection dependence of the emission blueshift and linewidth broadening of III-nitride quantum wells. We benchmark our calculations against electroluminescence (EL) measurements of a polar InGaN quantum-well device, and show that our calculation explains the experimentally observed injection dependence of the EL spectrum. In context of these results, we identify design strategies that minimize the wavelength shift and linewidth broadening of III-nitride emitters.

We self-consistently solved the Schrödinger and Poisson equations using nextnano++[32] and an in-house code, with parameters determined from first-principles density-functional theory (DFT) calculations.[31,33–36] We provide details of our calculations, which account for alloy disorder and carrier localization, in the Supplementary Material. We calculated the spontaneous-emission spectrum at first considering only band-filling effects in the disordered landscape of the quantum well, later shifting the spectrum energies to account for band-gap shift effects. We verified the validity of such a shift by checking that polarization-charge screening and plasma renormalization lead predominantly to a rigid shift of the bands (Figure S2 in the Supplementary Material). To calculate the band-gap shift, we solved the one-dimensional Schrödinger and Poisson equations, using an in-house code. We treated many-body exchange-correlation effects of the free carriers in the local-density approximation, using the Perdew-Wang parametrization[37] of the Monte-Carlo calculation by Ceperley and Alder.[38] This treatment of exchange and correlation accurately describes the experimentally measured[17] band-gap renormalization of bulk GaN (Figure S3 in the Supplementary Material). In the Supplementary Material, we provide details of the spectrum calculation and a discussion on the impact of alloy disorder on the numerical modelling of free-carrier screening and many-body renormalization, which explains our choice of calculating the band-gap shift effects with a one-dimensional rather than a three-dimensional solver.

To validate the accuracy of our calculations, we performed experimental measurements of the current-dependent EL spectrum of an InGaN LED packaged at Lumileds. These LEDs were designed so that practically all of the recombination occurs over a single quantum well, thus allowing us to determine the carrier density, which is needed to compare experiment with simulation. We discuss details of this "quasi-single-quantum-well" LED in the Supplementary Material. We performed electroluminescence measurements of the quasi-single-quantum-well LED under pulsed operation to minimize Joule heating, while ensuring that the time-averaged current density is only 1% of the peak current density. Our measurements exhibit both a current-dependent blueshift of the peak emission energy and broadening of the spectral linewidth (Figure 2(a)). The injection-dependent broadening is stronger on the high-energy side of the luminescence spectrum, which other groups have observed as well.[23–26,39] In order to compare our measurements with theory, we measured the recombination lifetime and carrier density using a previously developed small-signal RF technique,[40,41] in which we acquired and simultaneously fit the input impedance and modulation response to an equivalent circuit model of the LED to obtain the differential carrier lifetime. We then integrated the differential carrier lifetime to obtain the full carrier lifetime.[42,43] Figure 2(b) shows the recombination lifetime as a function of the current density; we also show the equivalent carrier density calculated from the relation, $J = e n_{2D}/\tau$, where $J$ is the current density, $n_{2D}$ is the two-dimensional carrier density and $\tau$ is the recombination lifetime. By measuring the recombination lifetime at various current densities, we converted the current dependence of the EL spectra to a carrier-density dependence, which is directly accessible in our calculations.

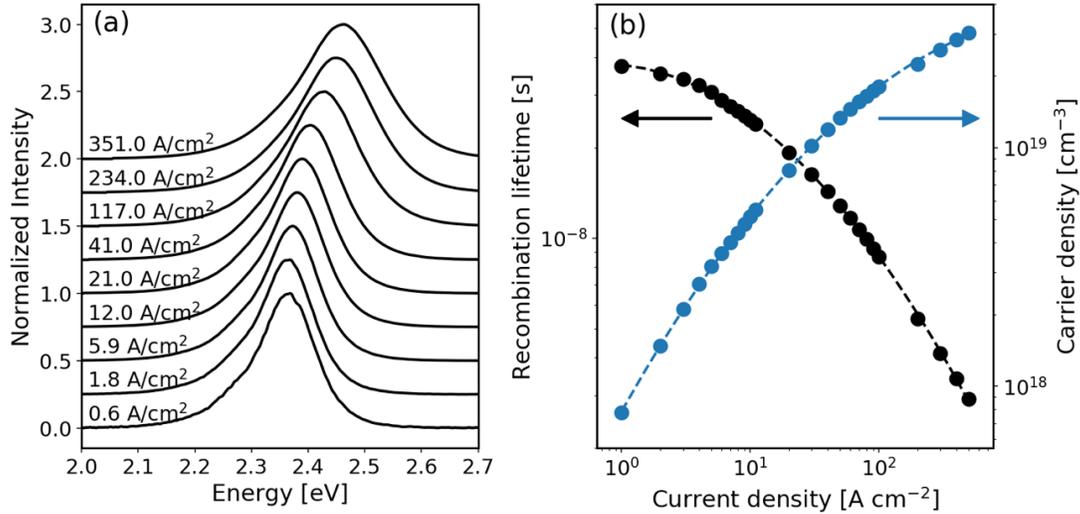

Figure 2. (a) Experimentally measured electroluminescence spectra of the InGaN quantum-well LED exhibiting a current-dependent blueshift and linewidth broadening. (b) Experimentally measured recombination lifetime (left axis) and the carrier density (right axis) calculated from the recombination lifetime, as a function of the injected current density.

Our modelling shows that we can accurately describe the carrier-density dependence of the peak-emission blueshift if we include the contributions of phase-space filling, polarization-charge screening, and many-body renormalization. In Figure 3, we show that our calculated carrier-density dependence of the peak-emission energy is in excellent agreement with experiment. We found that we needed to rigidly shift the band gap by $-0.4$ eV to quantitatively match the experimental gap, which suggests the presence of a systematic band-gap error in the modified $k \cdot p$ model.[44] We refer the reader to the Supplementary Material for a discussion on how we solved for the relative contribution of each effect to the net peak shift. We also find that for quantum wells with thicknesses of ~3 nm, the band-gap blueshift due to polarization-charge screening is compensated by a redshift due to plasma renormalization. We discuss the dependence of this cancellation on the quantum-well thickness in the Supplementary Material. Importantly, we show that polarization screening, phase-space filling, and plasma renormalization do not independently describe the shape of the carrier-density dependence curve. Therefore, our results show that it is crucial to accurately capture the contribution of all three effects to correctly model the emission spectra of InGaN emitters.

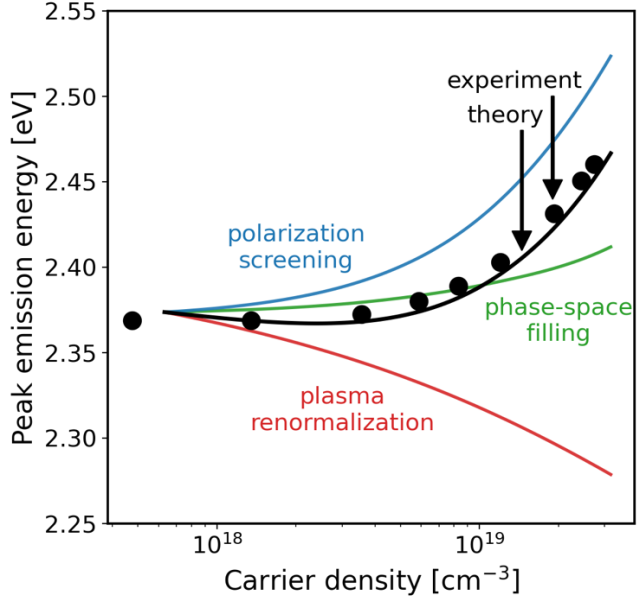

Figure 3. Theoretical carrier-density dependence of the peak emission energy of an InGaN quantum well (solid black curve) compared to experiment (scatter points). We show the relative contributions from polarization-charge screening (blue curve), phase-space filling (green curve) and plasma renormalization (red curve). There is excellent agreement between theory and experiment only if all three effects are included.

Furthermore, we find that phase-space filling of carriers in the disordered potential landscape of the InGaN quantum well accurately describes the experimentally measured linewidth broadening. In Figure 4(a), we show that our calculations of phase-space filling in the rigid-band approximation predict the relative increase of the full-width at half-maximum (FWHM) of the EL spectrum as a function of the carrier density. We report only the relative change to the FWHM rather than the exact value since only the former is physically meaningful due to the use of a constant energy-broadening parameter in calculating the joint density of states. As shown in Figure 4(b), a signature of phase-space filling is broadening of the high-energy luminescence tail, which is visible in the experimental EL spectrum of Figure 2(a) as well. According to the van-Roosbroeck-Shockley relation,[45] the low-energy tail of the luminescence spectrum corresponds to the shoulder of the joint density of states while the high-energy tail corresponds to the tail of the product of the electron and hole occupation functions. Since electrons are lighter than holes in the III-nitrides, the onset of hole degeneracy determines the onset of the broadening of the high-energy tail since both carriers need to be degenerate for phase-space filling to contribute to the peak wavelength blueshift and linewidth broadening. Since strongly localized carriers have smaller density of states than extended states, carrier localization exacerbates phase-space filling. However, localization is not a *requirement* for linewidth broadening, as previously conjectured,[23,25,27] since broadening of the Fermi tail is a general feature of degenerate-carrier statistics. Our observation that polarization-charge screening and plasma renormalization lead predominantly to a rigid shift of the bands (see Figure S2 of the Supplementary Material) further supports the argument that these two effects are less important than phase-space filling in explaining the linewidth broadening. We refer the reader to the Supplementary Material for further discussion on the impact of polarization fields and many-body effects on the linewidth broadening. Therefore, while the injection-dependence of the peak-emission energy is due to the interplay of various physical effects, the injection-dependent linewidth broadening is predominantly due to phase-space filling.

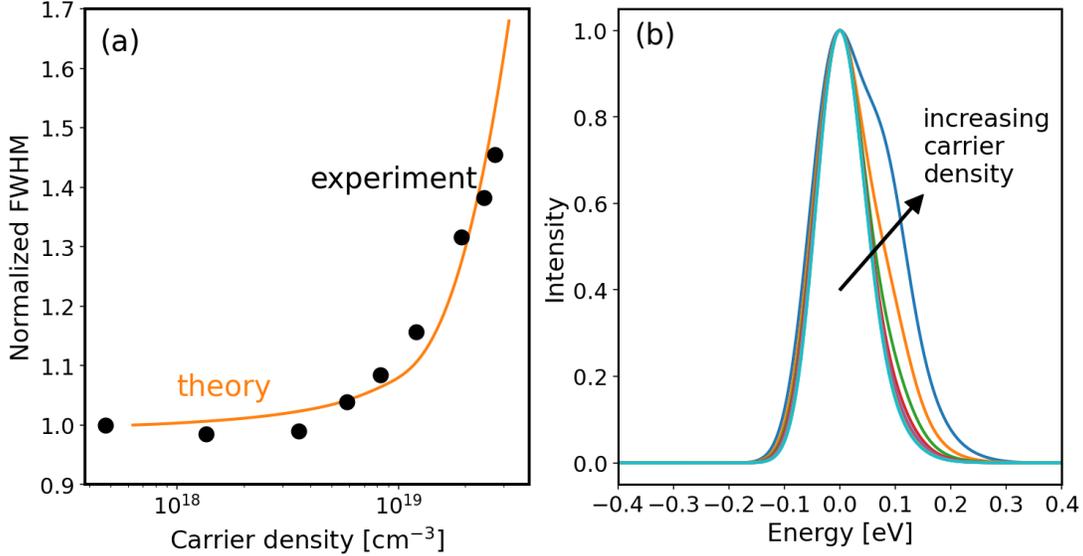

Figure 4. (a) Carrier-density dependence of the luminescence full-width at half-maximum due to phase-space filling of carriers in the disordered potential landscape of the InGaN quantum well. (b) Theoretical luminescence curve of a representative InGaN quantum well, with the peak-emission energy centered at zero. The signature of phase-space filling is broadening of the high-energy tail of the luminescence spectrum.

One important question that remains to be answered is why III-nitride LEDs grown on polar planes suffer from more severe injection-dependent linewidth broadening than III-phosphide and semipolar/non-polar III-nitride LEDs even though phase-space filling is a universal phenomenon that is present in all materials. The answer is simply that polar III-nitride LEDs operate at higher carrier densities due to their weaker oscillator strengths and correspondingly smaller radiative recombination (B) coefficients,[46] and are thus more susceptible to phase-space filling. In Figure 5, we show the carrier density required to operate 3 nm single-quantum-well LEDs at radiative current densities of 1A/cm$^2$, 50 A/cm$^2$, and 1000 A/cm$^2$ as a function of the B coefficient. We also show experimentally measured B coefficients for various (0001) polar[47] and (20$\bar{2}\bar{1}$) semipolar[48] LEDs. Polar LEDs have low B coefficients due to their strong polarization field, which separates electrons and holes to opposite sides of the quantum well and lowers the probability of recombination. The B coefficient of polar LEDs decreases with increasing emission wavelength (or indium content), therefore longer wavelength emitters undergo more severe injection-dependent spectral broadening. In contrast, semipolar LEDs have higher B coefficients due to their smaller polarization fields; consequently, they can operate at much lower carrier densities for a given current density. For this reason, semipolar LEDs exhibit less injection-dependent linewidth broadening than polar LEDs, a conclusion that is directly supported by optical measurements of semipolar LEDs in the literature.[49–52] The B coefficient of III-phosphide LEDs tend to be even higher than semipolar III-nitride LEDs, with typical B coefficients of the order $\sim$10$^{-10}$ cm$^3$ s$^{-1}$.[53] In fact, such high radiative recombination coefficients means that III-phosphide LEDs are more likely to experience stimulated emission before undergoing significant linewidth broadening, which may explain why luminescence broadening is typically not observed in the III-phosphide system. Our results also explain why some non-polar LEDs exhibit an (often small) injection-dependent blueshift and linewidth broadening despite the absence of a polarization field.[26,54,55] Because there is no quantum-confined Stark effect in non-polar LEDs, higher indium compositions are required to obtain a given wavelength. Carrier localization due to stronger alloy disorder reduces the density of states and lowers the B coefficient (if electrons and holes are not co-localized),[4,56] which makes phase-space filling important in non-polar LEDs. Although our analysis has been for InGaN LEDs, it applies equally well to AlGaN quantum-well LEDs, which also have strong polarization fields[57] and carriers localized by alloy disorder.[58] Hence, we have shown that recombination coefficients, and in particular the B coefficient, are important parameters that determine the likelihood of a device undergoing phase-space filling and injection-dependent linewidth broadening.

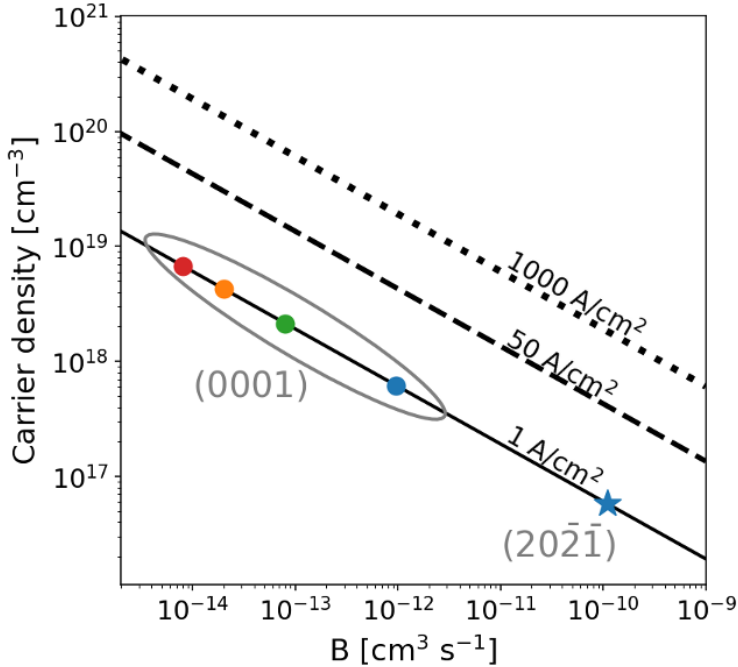

Figure 5. Effect of the B coefficient on the carrier density required to obtain a given radiative current density. The circles correspond to experimental B coefficients for polar (0001) LEDs measured by David et al. for blue (450 nm), green (535 nm), orange (600 nm), and red (645 nm) emitters.[47] The star is the experimental B coefficient measured by Monavarian et al. for a semi-polar ($20\bar{2}\bar{1}$) blue LED (430 nm).[48] LEDs with lower B coefficients are more susceptible to phase-space filling, and consequently to stronger spectral broadening, because they operate at higher carrier densities for a given current density.

Our results demonstrate that device designs that reduce the carrier density required to operate the device at a given current density reduce the injection-dependent blueshift and linewidth broadening. Improving the inter-well hole transport and spreading the number of carriers over more quantum wells enables the same light-power output for a lower carrier density. 3D engineering of the active region using V-pits has recently been shown to be a practical way of improving hole transport,[59] as evidenced by state-of-the-art multi-quantum-well LEDs fabricated with 3D V-pit engineering that show improved efficiency droop as well as smaller wavelength shift and linewidth broadening compared to LEDs with poor inter-well hole transport.[60] Designs that minimize the polarization field, e.g., semi-polar, non-polar, and thinner polar LEDs, minimize the injection-dependent wavelength blueshift because they allow the device to be operated at a lower carrier density for a given current density. Such designs simultaneously reduce the injection-dependent linewidth broadening and reduce efficiency droop, albeit at the expense of also requiring higher indium concentrations, which may inadvertently lead to a broader linewidth at low carrier density. In contrast, inefficient designs with more defects also operate at lower carrier densities for a given current density due to their higher non-radiative recombination rates, and thus exhibit less linewidth broadening. In general, it is important to identify the origin of small injection-dependent linewidth broadening, particularly in devices that are more susceptible to defects, e.g., micro-LEDs, as it can be a reflection of their high non-radiative recombination rate, which is highly undesirable. We highlight that the designs that minimize efficiency droop by reducing the operating carrier density of LEDs also lead to better color purity.

In summary, we have calculated the carrier-density dependence of the emission spectrum of InGaN LEDs. In contrast to the widely accepted hypothesis that the injection-dependent emission blueshift in III-nitride LEDs is primarily due to polarization-charge screening, we have shown that the emission shift depends on a complex interplay between polarization-charge screening, exchange-correlation effects, and phase-space filling of carriers in the disordered potential landscape of the quantum well. We have also shown that the injection-dependent linewidth broadening is caused primarily by phase-space filling, which is exceptionally prominent in polar III-nitride quantum wells due to their weaker oscillator strengths and lower radiative recombination coefficients. This emphasizes the innate connection between carrier dynamics and the current-dependent spectral characteristics of LEDs. Namely, emitters with poor transport and recombination dynamics offer poorer control over the injection-dependent color purity. Hence, designs that reduce the carrier density required to operate the LED at a given current density simultaneously reduce efficiency droop and improve the high-power color purity of III-nitride LEDs.

See the supplementary material for (1) details of our modified $k \cdot p$ calculations, (2) details of how we calculated the spontaneous emission spectrum, (3) details of the quasi-single-quantum-well InGaN LED on which we performed EL measurements, (4) a discussion on the impact of localization and alloy disorder on the modelling of free-carrier screening, (5) details of how we obtained the relative contributions of phase-space filling, polarization-charge screening, and plasma renormalization to the peak emission shift in Figure 3, (6) a discussion of the competition between polarization-charge screening and plasma renormalization in shifting the band gap, (7) a discussion of the effects of polarization-charge screening and plasma renormalization on the linewidth broadening, (8) a discussion of the merits and drawbacks of various designs that reduce the carrier density of LEDs at a given current density, (9) evidence that polarization-charge screening and plasma renormalization lead predominantly to a rigid shift of the bands (Figure S2), and (10) a comparison of the plasma renormalization of GaN calculated in the local-density approximation to experimental measurements of bulk GaN (Figure S3).

## Acknowledgements

We thank Siddharth Rajan for useful discussions. This project was funded by the U.S. Department of Energy, Office of Energy Efficiency & Renewable Energy, under award no. DE-EE0009163. Computational resources were provided by the National Energy Research Scientific Computing Center, a Department of Energy Office of Science User Facility, supported under Contract No. DEAC0205CH11231. N. Pant acknowledges the support of the Natural Sciences and Engineering Research Council of Canada Postgraduate Doctoral Scholarship.

(Supplementary Material)
# Origin of the injection-dependent emission blueshift and linewidth broadening of III-nitride light-emitting diodes

Nick Pant[1,2], Xuefeng Li[3], Elizabeth DeJong[3], Daniel Feezell[3], Rob Armitage[4], and Emmanouil Kioupakis[2]

[1]Applied Physics Program, University of Michigan, Ann Arbor, MI 48109, USA
[2]Department of Materials Science & Engineering, University of Michigan, Ann Arbor, MI 48109, USA
[3]Center for High Technology Materials, University of New Mexico, Albuquerque, NM 87106, USA
[4]Lumileds LLC, San Jose, CA 95131, USA

Corresponding author: nickpant@umich.edu

## 1. Details of our modified $\boldsymbol{k} \cdot \boldsymbol{p}$ Schrödinger-Poisson calculations

As input to our Schrödinger-Poisson calculations, we used elastic constants obtained in the local-density approximation[1] and improper polarization constants,[2] deformation potentials,[3] and band gaps and offsets[4] calculated with hybrid-functional DFT. To obtain room temperature values for the band gaps, we used empirical Varshni parameters,[5] although the temperature-dependent band-gap narrowing is very weak in the III-nitrides. We used the two-band effective-mass model for the conduction and valence bands, which is justified since we are interested in the band-edge optical properties.[6–8] We used $m_e^* = 0.19\ (\parallel), 0.21(\perp)$ and $m_h^* = 1.89$ for GaN, and $m_e^* = 0.07$ and $m_h^* = 1.81$ for InN, which are consistent with hybrid-functional[9] and many-body-perturbation-theory calculations.[10]

For our 3D calculations with nextnano++, we simulated thirty supercells of size 18 nm × 18 nm × 21 nm containing an InGaN quantum well with periodic boundaries, which is a valid approximation to the quantum well in an LED since the junction field is negligible if the device is fully turned on. To account for alloy disorder, we randomly assigned the composition in each grid site as either InN or GaN, and did not perform any further compositional averaging. We used a grid-size spacing of 0.3 nm in all directions, which corresponds to the intercation distance in (In)GaN. As input to our modelling, we used the out-of-plane composition profile of a commercial device, which we measured experimentally using energy-dispersive X-ray spectroscopy (EDS) and cross-validated with X-ray diffraction (XRD) measurements. Using this approach, we find that holes near the valence-band edge are localized within the plane due to alloy disorder, meanwhile electrons are extended within the plane.

For our 1D calculations with our in-house code, we self-consistently solved the one-dimensional Schrödinger and Poisson equations with a grid spacing of 0.01 nm. As mentioned in the main text, we accounted for many-body effects using the local-density approximation for the exchange-correlation potential. We screened the local-density exchange-correlation potential with the low-frequency dielectric constant $\epsilon_0$. We treated the electron and hole renormalization independently, in accordance with previous work on other semiconductors.[11–13] This approach allows us to self-consistently calculate the band-gap shift effects due to polarization-charge screening and many-body band-gap renormalization.

## 2. Calculation of the spontaneous-emission spectrum

We calculated the spontaneous-emission spectrum with the equation,

$$R_{sp}(\hbar\omega) = \frac{e^2 n_{r\omega}}{\hbar m_0 \epsilon_0 c^3 V} \frac{|p_{cv}|^2}{3} \sum_{n,m} f_n f_m \left| \int_V d^3 r \psi_n(r) \psi_m(r) \right|^2 \delta(\varepsilon_n - \varepsilon_m - \hbar\omega) \qquad (1)$$

where $\omega$ is the photon frequency, $n_r$ is the refractive index, $V$ is the recombination volume, $p_{cv}$ is the bulk interband momentum matrix element between the conduction and valence bands, $\psi_n$ and $\psi_m$ are electron and hole envelope functions, $f_n$ and $f_m$ are electron and hole occupation factors, and $\varepsilon_n$ and $\varepsilon_m$ are electron and hole energies. We approximated the delta function with a gaussian, and used a broadening parameter of 50 meV. We obtained the quasi-Fermi levels using the bisection method for root finding, assuming Fermi-Dirac statistics. To account for phase-space filling, we calculated the spontaneous-emission spectrum in the rigid-band approximation. We separately calculated the change to the band gap due to polarization-charge screening and many-body renormalization at the level of the local-density approximation and the virtual-crystal approximation. We then combined these two calculations to obtain the net carrier-density dependence of the peak-emission energy, as described in the main text, thus accounting for phase-space filling, polarization-charge screening, and many-body renormalization.

## 3. Details of the quasi-single-quantum-well LEDs

In state-of-the-art green LEDs, the growth of the active layers is optimized around V-defects in order to enable efficient hole injection into quantum wells farther away from the p-side of the device.[14] In such LEDs, recombination occurs in multiple quantum wells, resulting in improved EQE droop; however, this also gives rise to large uncertainty in estimating the carrier density, which is needed for comparison to theory. To avoid the uncertainty in carrier density, a quasi-single-quantum-well LED of simplified epitaxial design was the focus of our experimental study. This simplified device is representative of the quantum-well recombination dynamics in state-of-the-art LEDs, but not the inter-well carrier transport. The active region is comprised of three 3 nm quantum wells but practically all of the recombination occurs in the well closest to the p-type AlGaN electron blocking layer. This conclusion is supported by analysis of the measured angular distribution of the far-field radiation of unencapsulated planar LEDs[15] and also by comparing the device characteristics to those of an otherwise equivalent LED with the two wells closer to the n-side of the junction modified to emit blue instead of green by reducing their indium concentrations. The latter LED shows an obvious difference in photoluminescence spectra but its current-dependent electroluminescence characteristics (spectra, EQE, and forward voltage) are practically identical to those of the studied LED having three green wells. The epitaxial wafers were fabricated into LEDs using established manufacturing processes at Lumileds, and packaged into LUXEON C packages for testing.

## 4. Treatment of free-carrier screening and many-body effects in the presence of alloy disorder and carrier localization

Although the LDA works well for free-carrier plasmas in the virtual-crystal approximation, it cannot be faithfully applied to three-dimensional calculations with alloy disorder. As the plasma becomes more inhomogeneous, the LDA exchange becomes less effective in cancelling the spurious self-interaction of occupied carriers caused by the Hartree approximation.[16] Therefore, we have chosen to perform our calculations of polarization-charge screening and many-body renormalization in the virtual-crystal approximation, where the use of the LDA exchange-correlation is justified. Nevertheless, we do not expect the conclusions of our one-dimensional virtual-crystal calculations to change in the presence of carrier localization. Localized holes will screen the polarization charge more poorly compared to extended virtual-crystal states, however they will also contribute to a smaller band-gap renormalization due to reduced Coulomb matrix elements with other holes. Hence, we expect a cancellation of errors between polarization-charge screening and many-body renormalization in one-dimensional calculations, which justifies our virtual-crystal treatment of free-carrier screening. We note that this is an improvement over previous works that have neglected many-body exchange-correlation effects entirely.[6,8,17–19]

## 5. Calculation of the relative contributions to the peak shift in Figure 3 of the main text

In Figure 3 of the main text, we calculated the relative contribution of phase-space filling in the rigid-band approximation because polarization-charge screening and plasma renormalization can be treated, to first order, as a rigid shift of the bands, as discussed earlier in the text. We obtained the relative contribution of polarization-charge screening by calculating the band-gap shift in a one-dimensional calculation at the level of the mean-field Hartree approximation. Finally, we obtained the relative contribution of plasma renormalization by taking the difference of the band-gap shift between the Hartree and local-density approximations.

## 6. Competition between polarization-charge screening and plasma renormalization in shifting the band gap

The blueshift of the band gap due to polarization-charge screening is compensated by a redshift of the band gap due to plasma renormalization.

We demonstrate that the quantum-well thickness influences how polarization-charge screening and many-body effects compete in shifting the band gap. In the range of carrier densities relevant for LED operation, the band gap remains approximately independent of the carrier density due to a cancellation of the blueshift due to polarization-charge screening by the redshift due to plasma renormalization, as shown in Figure S1(a). The cancellation of these two effects depends on the quantum-well thickness. As a point of comparison, we consider the carrier-density range between $3\times10^{10}$ cm$^{-2}$ and $3\times10^{12}$ cm$^{-2}$. In Figure S1(b), we show that there is an approximate cancellation of the two effects for 3 nm quantum wells, which is the thickness typically used in commercial LEDs, thus the blueshift in this range is predominantly due to band-filling effects rather than band-gap shift effects. Overall, thicker wells experience a net band-gap blueshift while thinner wells experience a net redshift, thus reflecting the challenge in fabricating long-wavelength emitters based on thick polar quantum wells.

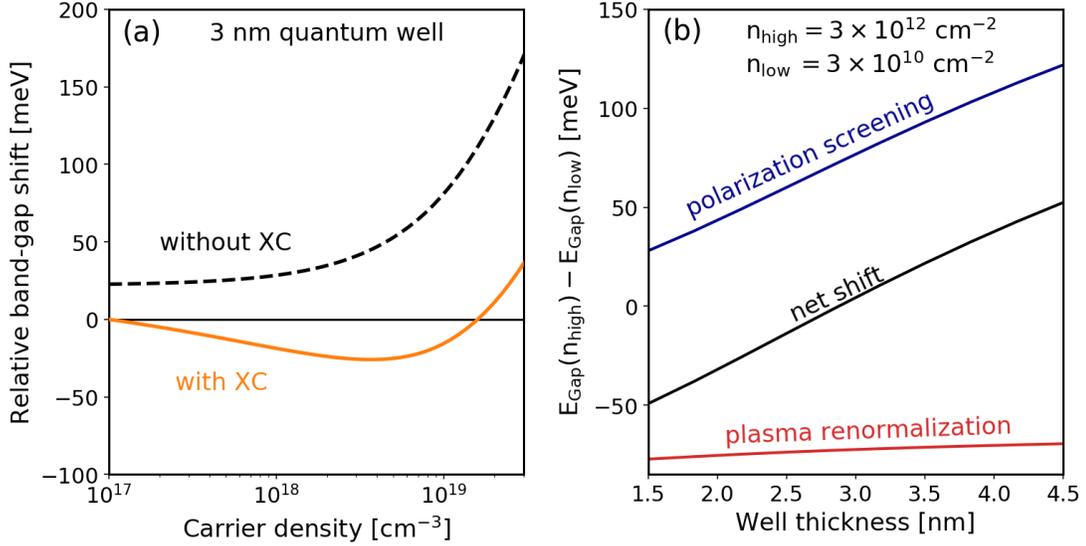

Figure S1. (a) Relative band-gap shift of a 3 nm InGaN quantum well, compared to the band gap at a carrier density of $10^{17}$ cm$^{-3}$, with (solid curve) and without (dashed curve) exchange-correlation (XC) effects, showing the importance of including many-body effects in calculations to describe the band-gap shift. (b) The band-gap shift of green-emitting quantum wells between carrier densities of $n_{low} = 3\times10^{10}$ cm$^{-2}$ and $n_{high} = 3\times10^{12}$ cm$^{-2}$, as a function of the quantum-well thickness. There is virtually no net band-gap

shift from $n_{low}$ to $n_{high}$ for 3 nm quantum wells due to a fortuitous cancellation between polarization screening and plasma renormalization.

## 6. Contribution of effects other than phase-space filling to the linewidth broadening

We argue that the injection-dependent linewidth broadening must be predominantly determined by phase-space filling, rather than polarization-charge screening or many-body renormalization. Indeed, polarization-charge screening leads to the removal of the quantum-confined Stark effect, which lifts the level repulsion between states that were mixed by the electric field. This increases the density of states near the band edge, thus shrinking the region of phase-space that carriers occupy. Such an effect would lead to linewidth narrowing, which is qualitatively inconsistent with the experimentally observed broadening, therefore polarization-charge screening can be ruled out as a source of the broadening. Polarization fields have an additional *second-order* effect on the linewidth broadening since polarization fields modulate the B coefficient, which controls the carrier density required to operate the LED at a given current density ($J = Bn^2$). Weaker polarization fields, e.g., due to screening, increase the B coefficient, thus reducing the carrier density at a given current density, thereby lessening phase-space filling. This effect is still relatively small compared to first-order phase-space filling effects at the carrier densities that we have considered, and indeed would *reduce* the relative linewidth broadening as the carrier density increases. Moreover, in the carrier densities relevant for LED operation, many-body renormalization of the band structure can be treated, to first order, as a rigid shift of the band edges,[12] thus it does not contribute strongly to linewidth broadening either. This is supported by our observation that the net result of polarization-charge screening and plasma renormalization is approximately a rigid shift of the bands (see Figure S1).

## 7. Merits and drawbacks of designs that reduce the carrier density of LEDs at a given current density

Some authors have suggested that adding indium to the barriers may enable better hole transport in multi-quantum-well LEDs.[20] Although such designs work well for blue emitters, it is unclear whether they work for green and longer wavelength emitters since adding indium to the barrier increases the emission energy. A similar argument applies for doping the barriers to minimize the polarization fields in polar multi-quantum-well structures.[21] As discussed in the main text, thinner polar quantum wells are more desirable than thicker polar quantum wells due to their higher oscillator strengths and larger B coefficients, which allows them to operate at lower carrier densities for a given current density. However, higher indium concentrations are needed in thinner quantum wells to obtain a given wavelength, which may lead to a broader linewidth at low carrier density due to stronger alloy disorder. Moreover, polar quantum wells with thickness greater than 3 nm experience an additional net blueshift of the band gap for high carrier densities, due to an incomplete cancellation of the polarization-charge blueshift by the many-body redshift, therefore they are especially undesirable for applications that require good color purity. Conversely, thicker non-polar quantum wells are more desirable than thinner non-polar quantum wells since the current density achievable for a given carrier density scales linearly with the thickness if the B coefficient is held constant. Overall, any strategy that reduces the B coefficient by reducing the polarization field will inadvertently increase the indium concentration required to obtain a given wavelength. Therefore, such strategies will require a careful tradeoff between an increase of the B coefficient due to minimizing polarization fields versus stronger carrier localization and material degradation due to increasing indium concentrations.

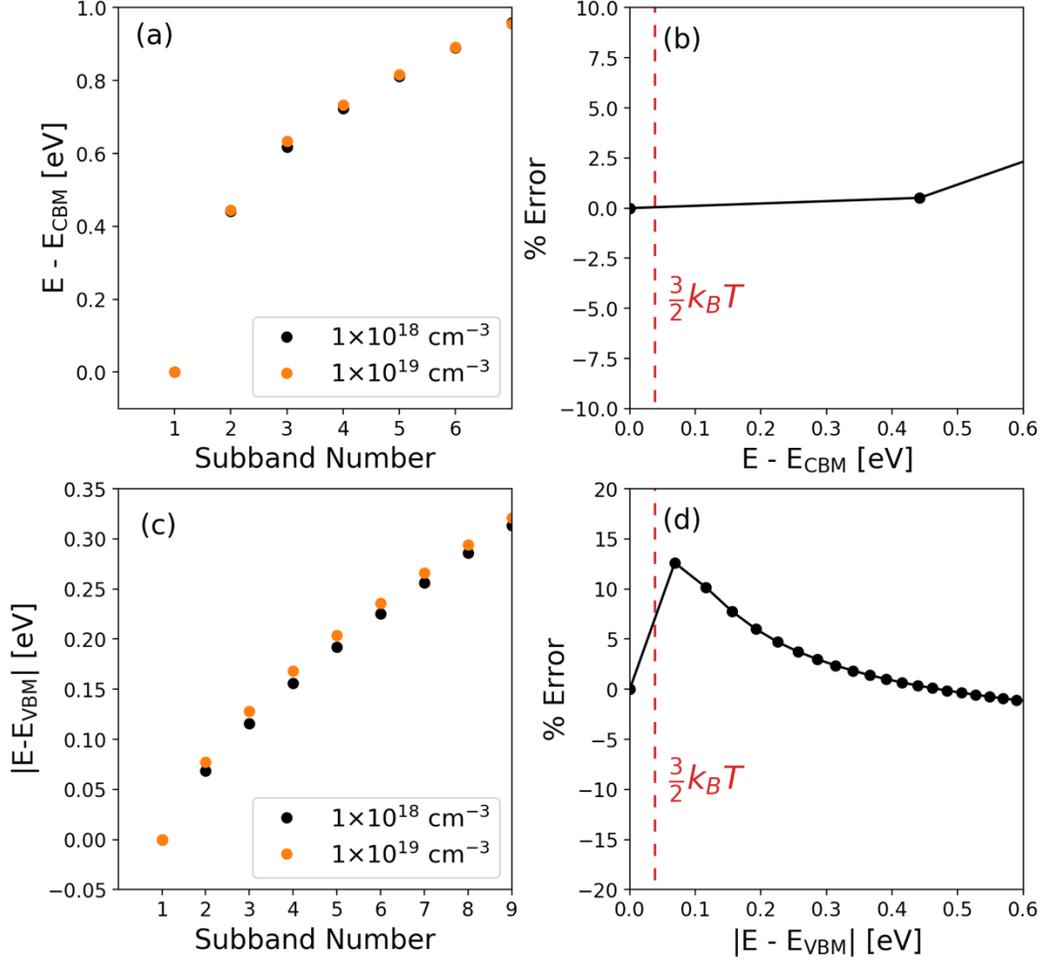

Figure S2. Evidence that polarization-charge screening and plasma renormalization lead predominantly to a rigid shift of the bands in the carrier-density range of interest for LED operation. Panel (a) compares the electron energy of the subbands in an InGaN quantum well with carrier densities of $10^{18}$ cm$^{-3}$ and $10^{19}$ cm$^{-3}$, and panel (b) shows the relative error accrued by assuming the conduction band is rigidly shifted due to screening effects. The error in the conduction band accrued by assuming a rigid shift of the bands is negligible. Panel (c) compares the hole energy of the subbands in an InGaN quantum well with carrier densities of $10^{18}$ cm$^{-3}$ and $10^{19}$ cm$^{-3}$, and panel (d) shows the relative error accrued by assuming the valence band is rigidly shifted due to screening effects. The error in the valence band accrued by assuming a rigid shift of the bands is small; the largest error is for the first excited subband, however the error is small (less than 15%), which is further diminished by the fact that the thermal occupation of this subband is small.

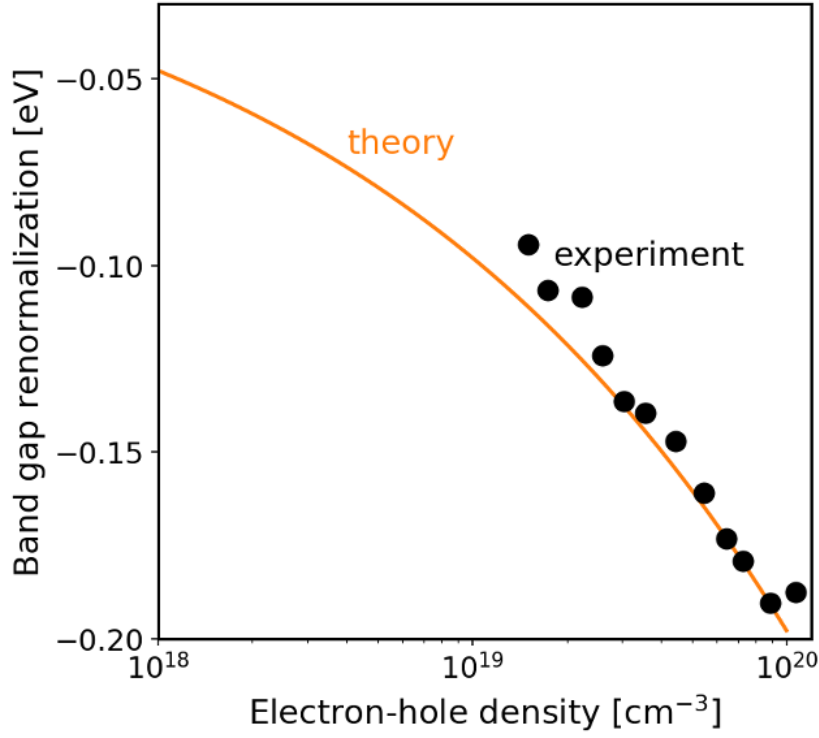

Figure S3. Theoretical band-gap renormalization by free carriers due to many-body exchange-correlation effects in bulk GaN (solid curve), compared to experimental measurements (scatter points) by Nagai et al.[22]